# Aggregate Analytic Window Query over Spatial Data

**Xing Shi and Chao Wang**

Guangdong University of Technology, Guangzhou, Guangdong 510006, China
North China University of Technology, Beijing 100144, China

**Abstract.** Analytic window query is a commonly used query in relational databases. It answers the aggregations of data over a sliding window. For example, to get the average prices of a stock for each day. However, it is not supported in the spatial databases. Because the spatial data are not in a one-dimension space, there is no straightforward way to extend the original analytic window query to spatial databases. But these queries are useful and meaningful. For example, to find the average number of visits for all the POIs in the circle with a fixed radius for each POI as the center. In this paper, we define the aggregate analytic window query over spatial data and propose algorithms for grid index and tree-index. We also analyze the complexity of the algorithms to prove they are efficient and practical.

## 1. **Introducion**

Aggregate analytic query is a widely used query in relational databases. It sorts the data records by a one-dimensional attribute and gets the aggregations of a sliding window. It returns a single result for each row, which is different from an aggregate function that returns a single result for an entire group of rows. There is an OVER clause in such queries, which defines a window of rows around the row to be evaluated. However, there is no easy extension of analytic query from one dimension to two dimensions, so the spatial aggregate analytic query is not easily supported. The spatial aggregate analytic query is meaningful and practical in the data analysis and visualization of spatial data. For example, show the total number of visits of POIs (point-of-interest) in each range (each range is defined by a fixed radius, and the center of the range circle is each POI). In this paper, we define the aggregate analytic window query over spatial data and propose techniques to answer the query in spatial databases with different indexes, such as grids and quadtree. Different various types of aggregate analytic queries may have different aggregate functions. They can be divided in two groups, const-memory-used functions such as COUNT, AVG, SUM and non-const-memory-used functions such as MAX, MIN. We only discuss const-memory-used functions in this paper. We analyze the complexity of the algorithms and prove their efficiency.

## 2. **Model**

### 2.1. kNN Window and Range Window

#### *2.1.1. kNN Window*

An aggregate analytic query over the relational databases is as the following examples. Query 1 shows an example of the row-based window.

SELECT StudentID, Score, AVG(Score) OVER (ROW BETWEEN 2 PRECEDING AND 1 FOLLOWING) FROM student_list;

Here it returns the average scores of each student's previous two students, himself and one following student and in the default order of the current table.

**Table 1.** Student List.

| StudentID | Name | Score |
|---|---|---|
| 000000001 | David | 90 |
| 000000002 | Justin | 70 |
| 000000003 | Alice | 89 |
| 000000004 | Bob | 80 |
| 000000005 | Lucy | 81 |
| 000000006 | Lily | 75 |
| 000000007 | Ray | 86 |

**Table 2.** The result of Query 1.

| StudentID | Score | AVG(Score) |
|---|---|---|
| 000000001 | 90 | (90+70)/2 |
| 000000002 | 70 | (90+70+89)/3 |
| 000000003 | 89 | (90+70+89+80)/4 |
| 000000004 | 80 | (70+89+80+81)/4 |
| 000000005 | 81 | (89+80+81+75)/4 |
| 000000006 | 75 | (80+81+75+86)/4 |
| 000000007 | 86 | (81+75+86)/3 |

We extend the definition of the row based window to a similar concept in the n-dimensional space, the kNN window. Because there is no preceding and following in the n-dimensional space, we don't consider the direction of the neighbor. k will be given as a parameter like the (x, y) in the over clause "ROW BETWEEN x PRECEDING AND y FOLLOWING". The syntax of the spatial aggregate analytic query is defined as,

```
analytic_function_name ( [ argument_list ] ) OVER over_clause

over_clause:
  { named_window | ( [ window_specification ] ) }

window_specification:
  [ named_window ]
  [ window_frame_clause ]

window_frame_clause:
  { k NEAREST NEIGHBOUR ON location }
```

k is the given count of the nearest neighbours and location is a (latitude, longitude) type attribute.

For example, if we want to get the total number of the visits of k nearest neighbours and itself of all the POIs, the query can be,

SELECT poi_id, location, SUM(number_of_visits) OVER (k NEAREST NEIGHBOUR ON location) FROM poi_data;

#### *2.1.2. Range Window*

The range based window is bounded by the value of the current row and the range given. We take Query 2 as an example,

SELECT StudentID, Score, AVG(Score) OVER (ORDER BY Score RANGE BETWEEN 2 PRECEDING AND 1 FOLLOWING) FROM student_list;

The first step is to sort the table rows by Score, the result is Table. 3. The second step is to compute the average based on the value of each row and the given range, as Table. 4.

**Table 3.** Sorted by Score.

| StudentID | Score |
|---|---|
| 000000002 | 70 |
| 000000006 | 75 |
| 000000004 | 80 |
| 000000005 | 81 |
| 000000007 | 86 |
| 000000003 | 89 |
| 000000001 | 90 |

**Table 4.** The result of Query 2.

| StudentID | Score | AVG(Score) |
|---|---|---|
| 000000002 | 70 | 70 |
| 000000006 | 75 | 75 |
| 000000004 | 80 | (80+81)/2 |
| 000000005 | 81 | (80+81)/2 |
| 000000007 | 86 | 86 |
| 000000003 | 89 | (89+90)/2 |
| 000000001 | 90 | (89+90)/2 |

Similar to the kNN window, we extend the definition of the range based window to a similar concept in the n-dimensional space. Because there is no preceding and following in the n-dimensional space, we don't consider the direction of the range. A fixed radius will be given as a parameter like the (x, y) in the over clause "RANGE BETWEEN x PRECEDING AND y FOLLOWING". The syntax of the spatial aggregate analytic query is defined as,

```
analytic_function_name ( [ argument_list ] ) OVER over_clause

over_clause:
  { named_window | ( [ window_specification ] ) }

window_specification:
  [ named_window ]
  [ window_frame_clause ]

window_frame_clause:
  { RADIUS r ON location }
```

r is the given radius of the circle and location is a (latitude, longitude) type attribute.
For example, if we want to get the total number of the visits of neighbors whose distance is less than r and itself of all the POIs, the query can be,

SELECT poi_id, location, SUM(number_of_visits) OVER (RADIUS r ON location) FROM poi_data;

### 2.2. *Aggregation Functions*

Most aggregation functions can be also used as the analytic functions. Some of them only need const memory to remember the state of the current window, the others need linear memory space. For the const-memory-used analytic functions, the aggregation over all the rows/points in the current window does not have to be computed from zero. Only the rows/points newly entering/leaving the window need to be considered. The state in the memory is updated from the previous window by the new rows/points and the new aggregate can be computed based on the state. For example, for a sequence of points 1, 2, 3, 4, 5. Assuming the current moving window contains three items, 1+2+3+4. When we are trying to get the result of 2+3+4+5, we only need to add (5-1) to the previous state, rather than recomputing it, because only 5 and 1 are the rows/points newly entering/leaving the window.

Here we discuss them one by one and show which of them are the ones focused in this paper. Here a list

*ANY_VALUE, ARRAY_AGG, AVG, CORR, COUNT, COUNTIF, COVAR_POP, COVAR_SAMP, MAX, MIN, STDDEV_POP, STDDEV_SAMP, STRING_AGG, SUM, VAR_POP, VAR_SAMP*

The update method works for all of them except MIN/MAX, so they are const-memory-used functions, while not for MIN/MAX, we know that they are non-const-memory-used functions. In this paper, we only provide solutions for the const-memory-used functions.

## 3. **A scan algorithm by the grid index**

We propose a scan algorithm for spatial data with grid index first. For the range window, we can start from the grid with the minimal latitude and longitude. First we get the point with the minimal latitude or the minimal longitude, and issue a range query with radius r. And then find the next four points with minimal |latitude - latitude'| and |longitude - longitude'| for the four directions.

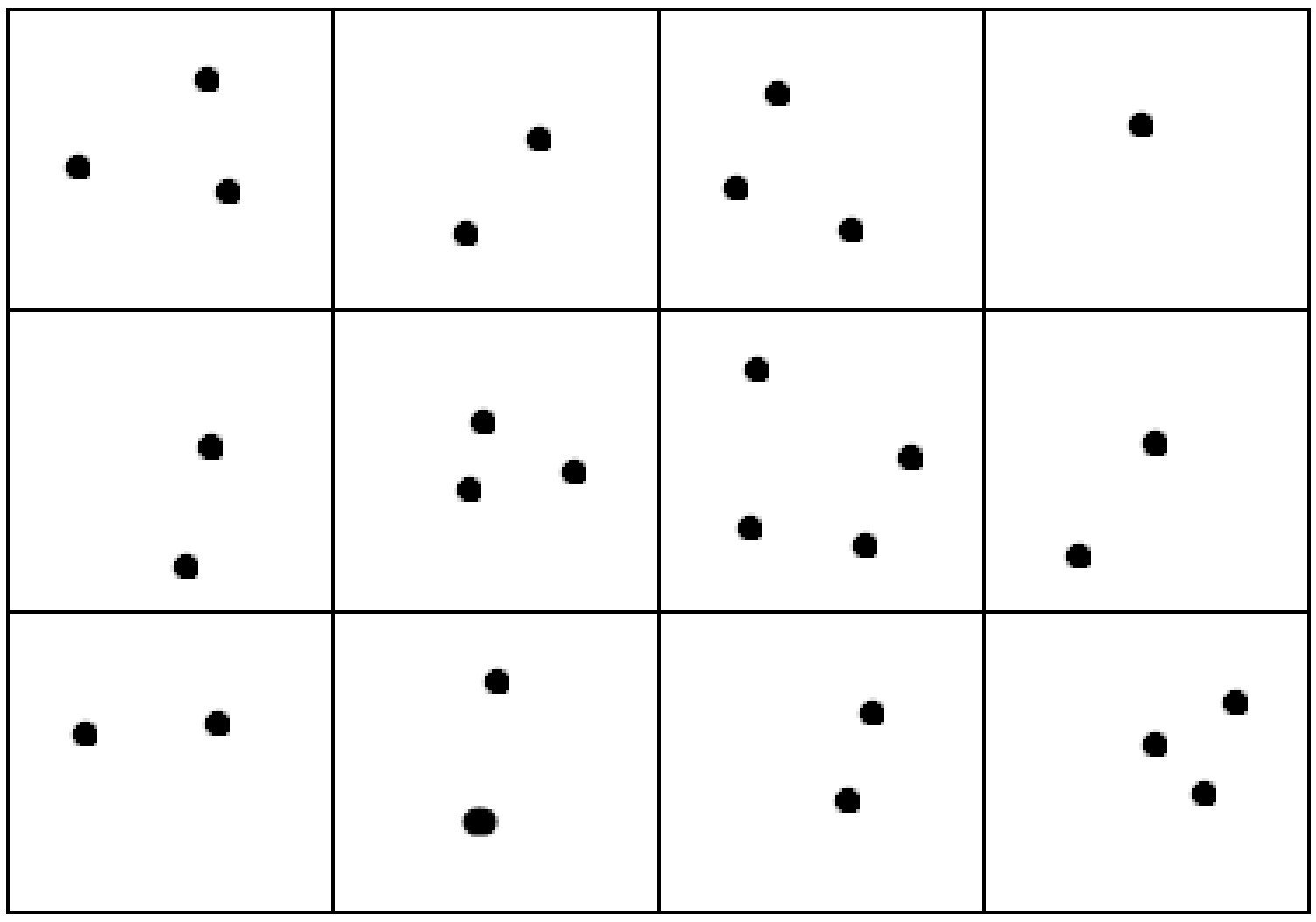

Figure 1. Spatial data with grid index.

## 4. **A scan algorithm by the quadtree index**

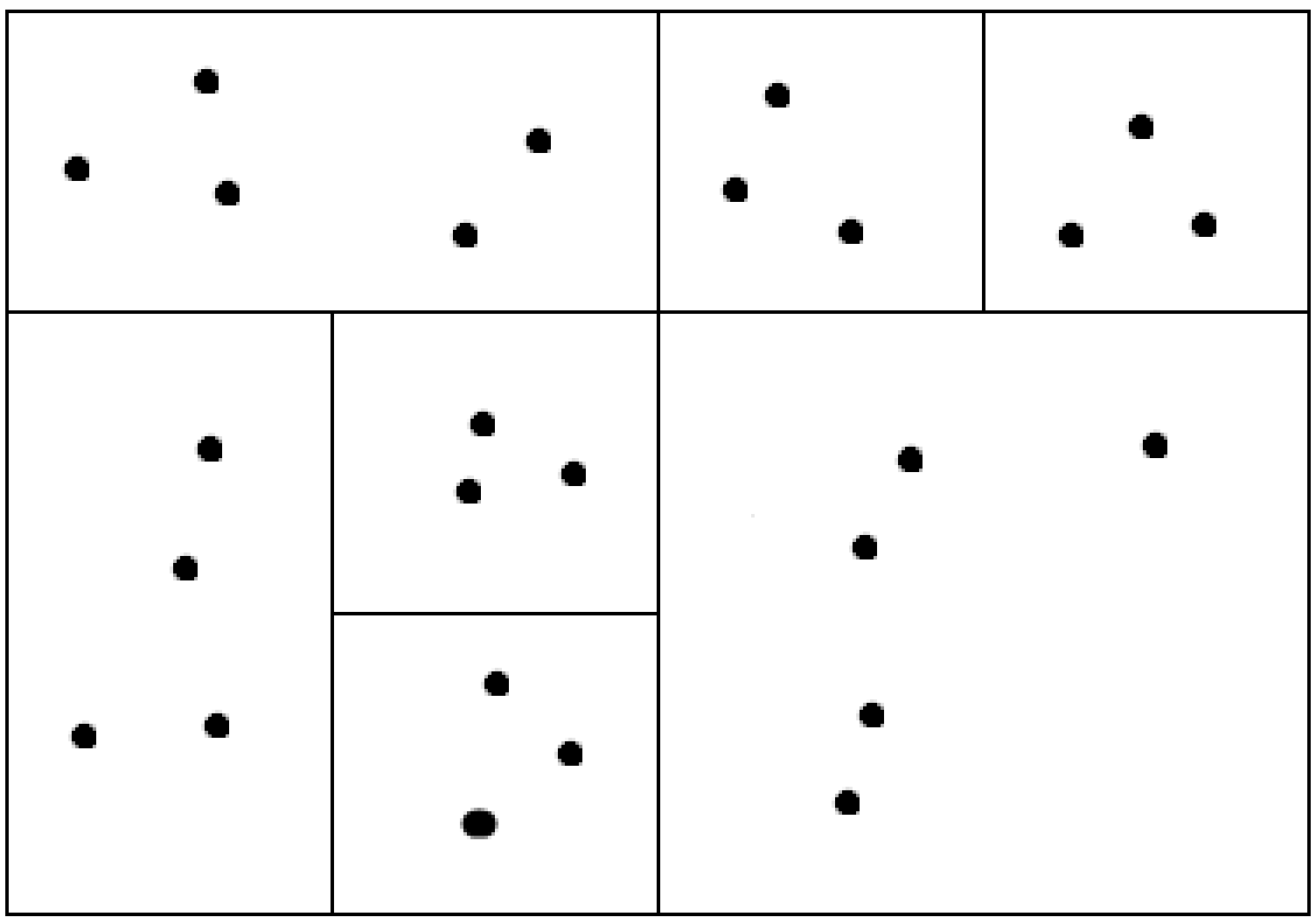

Figure 2. Spatial data with quadtree index.

## 5. **Related Work**

The research of [8] answers the aggregate estimations over the spatial data. It uses the area of the voronoi cell as the probability to make the estimation of the aggregations. [2] has a clear definition of the analytic functions for the relational databases. [10] applies the density based clustering over the spatial data, whose algorithms is a scan on the spatial data by a Hilbert curve. [11, 12] are trying to answer aggregate queries over the road networks, which includes the spatial data and the graph data.

## 6. **References**

[1] S. Agarwal, R. Agrawal, P. Deshpande, A. Gupta, J. F. Naughton, R. Ramakrishnan, and S. Sarawagi. On the computation of multidimensional aggregates. In VLDB, pages 506 – 521, 1996.

[2] S. Bellamkonda, T. Bozkaya, B. Ghosh, A. Gupta, J. Haydu, S. Subramanian, and A. Witkowski. Analytic functions in Oracle 8i. Technical report, 2000. http://tinyurl.com/3pcbsmq.

[3] D. Chatziantoniou and K. A. Ross. Querying multiple features of groups in relational databases. In VLDB, pages 295–306, 1996.

[4] Z. Chen and V. Narasayya. Efficient computation of multiple group by queries. In SIGMOD, pages 263–274, 2005.

[5] T. Neumann and G. Moerkotte. A combined framework for grouping and order optimization. In VLDB, pages 960–971, 2004.

[6] D. Simmen, E. Shekita, and T. Malkemus. Fundamental techniques for order optimization. In SIGMOD, pages 57–67, 1996.

[7] X. Wang and M. Cherniack. Avoiding sorting and grouping in processing queries. In VLDB, pages 826–837, 2003.

[8] Liu, Weimo, et al. "Aggregate estimations over location based services." arXiv preprint arXiv:1505.02441 (2015).

[9] Cao Y, Chan CY, Li J, Tan KL. Optimization of Analytic Window Functions. Proceedings of the VLDB Endowment. 2012;5(11).

[10] Rahman, Md Farhadur, et al. "Density based clustering over location based services." 2017 IEEE 33rd International Conference on Data Engineering (ICDE). IEEE, 2017.
[11] Sun, Weiwei, et al. "Merged aggregate nearest neighbor query processing in road networks." Proceedings of the 22nd ACM international conference on Information & Knowledge Management. 2013.
[12] Sun, Weiwei, et al. "Fast optimal aggregate point search for a merged set on road networks." Information Sciences 310 (2015): 52-68.